%% file: main.tex
\documentclass[review]{elsarticle}
\usepackage[margin=1in]{geometry}

\usepackage{lineno,hyperref}
\modulolinenumbers[5]

\usepackage{setspace}
\doublespace 

\journal{Journal of Sound and Vibration}

\usepackage{siunitx}
\usepackage{float}
\usepackage{amsmath}
\usepackage{amssymb}
\usepackage{textcomp}
\usepackage{subfigure}
\usepackage{graphicx}
\usepackage[colorinlistoftodos]{todonotes}

\begin{document}

\begin{frontmatter}

\title{Stress-optimized inertial amplified metastructure with opposite chirality for vibration attenuation}

\author[mainaddress]{Rachele Zaccherini\corref{mycorrespondingauthor}}
\cortext[mycorrespondingauthor]{Corresponding author at: Department of Civil, Environmental and Geomatic Engineering, ETH Z\"urich, 8093 Z\"urich, Switzerland.}
\ead{zaccherini@ibk.baug.ethz.ch}

\author[mainaddress]{Andrea Colombi}
\author[secondaryaddress]{Antonio Palermo}
\author[mainaddress]{Henrik R. Thomsen}
\author[mainaddress]{Eleni N. Chatzi}

\address[mainaddress]{Department of Civil, Environmental and Geomatic Engineering, ETH Z\"urich, 8093 Z\"urich, Switzerland}
\address[secondaryaddress]{Department of Civil, Chemical, Environmental and Materials Engineering - DICAM, University of Bologna, 40136 Bologna, Italy}

\begin{abstract}
In this work, we investigate the dynamics and attenuation properties of a one-dimensional inertial amplified lattice with opposite chirality. The unit cell of the structure consists of a hollow-square plate connected to a ring through arch-like ligaments. The peculiar geometry and orientation of the links allow for coupling the axial and the torsional motion of the lattice, thus amplifying the inertia of the system. We develop both simplified analytical and numerical models of the building block to derive the complex dispersion relation of the infinite lattice. The structure supports a frequency-tailorable attenuation zone, whose lower bound is controlled by the second coupled axial-torsional mode. Laboratory measurements of the transmission spectrum on a 3D printed sample match very well with the analytical and numerical predictions, confirming the wide-band filtering properties of this lattice. We complete our investigation by developing and solving a constrained optimization model to obtain the optimized geometric parameters of the unit cell that minimize the bandgap opening frequency and, at the same time, fulfill structural requirements. In particular, the internal stresses induced by the self-weight of the structure are kept to a low by virtue of the employed design, with the aim to prevent plastic deformations and failure. The inertial amplification mechanism, proposed and investigated in this work, offers an efficient variant for the efficient design of materials and structures for vibration mitigation and shock protection.
\end{abstract}

\begin{keyword}
Inertial amplification mechanism\sep Mechanical metamaterial\sep Vibration attenuation\sep Optimization
\end{keyword}

\end{frontmatter}


\section{Introduction}
Architected metamaterials are artificially-engineered materials comprised of periodically arranged building blocks, whose geometry, orientation, and size are ad hoc designed to attain unique properties, unachievable in conventional solids \citep{Schaedler,Bertoldi}. In addition to light weight \citep{Jacobsen,Spadaccini}, negative Poisson’s ratio \citep{Babaee}, heat dissipation \citep{Jia}, and high-energy absorption \citep{Lee} materials, there exist metamaterials that comprise unconventional vibrational characteristics \citep{Scarpa,Chen}. These mechanical metamaterials are, nowadays, well-known for their capacity to attenuate \citep{AndreaForest,Metabarrier} and trap \citep{AndreaTrap,GregTrap} elastic waves over certain frequency bands, referred to as bandgaps. The design of metamaterials equipped with wide bandgaps and good isolation characteristics within the generated stop-bands offers benefits for various applications, including wave filtering \citep{Rupp}, waveguiding \citep{Sigmund,Khelif}, energy harvesting \citep{Ponti}, and sound and vibration isolation \citep{Alessandro,Zaccherini1,Zaccherini2}. Bandgaps are commonly generated via two different mechanisms, namely, Bragg scattering \citep{Sala,Vasseur} and local resonances \citep{Liu,Goffaux}. Bragg-induced bandgaps appear due to destructive interference between incident and scattered waves, the latter generated as a result of the metastructure periodicity, when the wavelength of the propagating wave is comparable to the structural periodicity. Instead, resonance-induced bandgaps are generated by the local resonances of scatterers embedded inside the metastructure matrix \citep{Liu}. In this case, it is the natural frequency of each resonator within the unit cell, and not the structural periodicity, which controls the onset of the bandgap.

One of the challenges in designing metamaterials for vibration attenuation is to obtain wide bandgaps at a low-frequency range, which is the desideratum for numerous applications, particularly those dealing with mechanical wave propagation and attenuation. The central frequency of the lowest Bragg-induced bandgap is of the order of the wave speed (longitudinal or transverse) of the medium divided by the lattice constant \citep{Sheng}. Therefore, to achieve a low frequency bandgap, either high density/low modulus materials or large lattice constants are required. Likewise, to obtain a wide resonance-induced bandgap at a low-frequency range, heavy and cumbersome resonant masses must be employed.

Recently, architected structures with inertial amplification mechanisms have been proposed to overcome the shortcomings of the standard periodic structures at low frequencies \citep{Acar,Taniker}. By amplifying the effective inertia of the structure via embedded amplification mechanisms, the bandgap opening frequency is decreased. As a result, lower and broader bandgaps are achieved, while keeping the metastructure lightweight.

A common investigated inertial amplification mechanism pertains to the arrangement of a small mass that is connected to two inclined moment-free rigid links \citep{Kikuchi}. When the inclination angle of the latter is small, the displacement of the small mass is amplified and the inertia of the system is increased. This mechanism has been used as a backbone structural component in discrete systems \citep{Kikuchi,Hulbert,Settimi}, but has been also incorporated into host structures, such as rods \citep{Bilal}, beams \citep{Li}, or mechanical oscillators \citep{Zeighami}. An alternative method to amplify the inertia of the system employs chiral connections to couple the axial and rotational motion of the metastructure mass elements \citep{Delpero,Krushynska,Orta}. The introduction of the rotational degree of freedom (DOF) amplifies the overall inertia of the metastructure, without increasing the total mass of the system. Recent studies \citep{Wang,Bergamini} demonstrate that, besides the exploitation of the concept of chirality, the relative orientation of the links strongly affects the nature of the coupling between the mass elements, considerably enlarging the attenuation bandwidth of the metastructure. To enhance structural flexibility and enable a good compression-torsion coupling, these links are often designed with thin geometries. As a result, these structures may feature high stress concentrations in critical locations, which can in turn compromise their functionality and structural safety. Limiting the internal stresses to prevent plastic deformations and static failure is an essential task during the structural design process. This is particularly true when these structures are used as load bearing connectors. However, the previous designs do not account for such stress limitations.

In this work, we exploit the concept of opposite chirality to design an inertial amplified structure for the mitigation of low-frequency mechanical vibrations, limiting the internal stresses induced due to self-weight to ensure a safe and durable design. The building block consists of a hollow-square plate connected to a ring through four arch-like ligaments. The curved shape of the ligaments promotes the coupling between axial and torsional motion of the structure, thereby increasing the effective inertia of the system. We develop an analytical lumped parameter model and a finite element model to obtain the complex dispersion diagram of the infinite structure. To investigate how the linear elastic wave propagation in the finite lattice is affected upon variation of the number of unit cells composing the lattice, we perform multiple frequency domain analysis. Next, we experimentally validate the inertial amplification mechanism. In particular, we fabricate a finite-sized lattice composed of two unit cells by selective laser sintering in Nylon 12 and test vibration mitigation capabilities via time-transient analysis. Finally, we optimize the design parameters of the unit cell by nonlinear minimization techniques. We perform constrained optimization studies to determine the structure's optimal design parameters that allow for achieving the lowest-frequency bandgap, while fulfilling given stress design conditions. In particular, we restrict the internal stresses due to self-weight to prescribed maximum admissible values, to ensure a safe design.

\section{Metastructure design}
\begin{figure}[H]
\centering
\includegraphics[width=1\textwidth]
{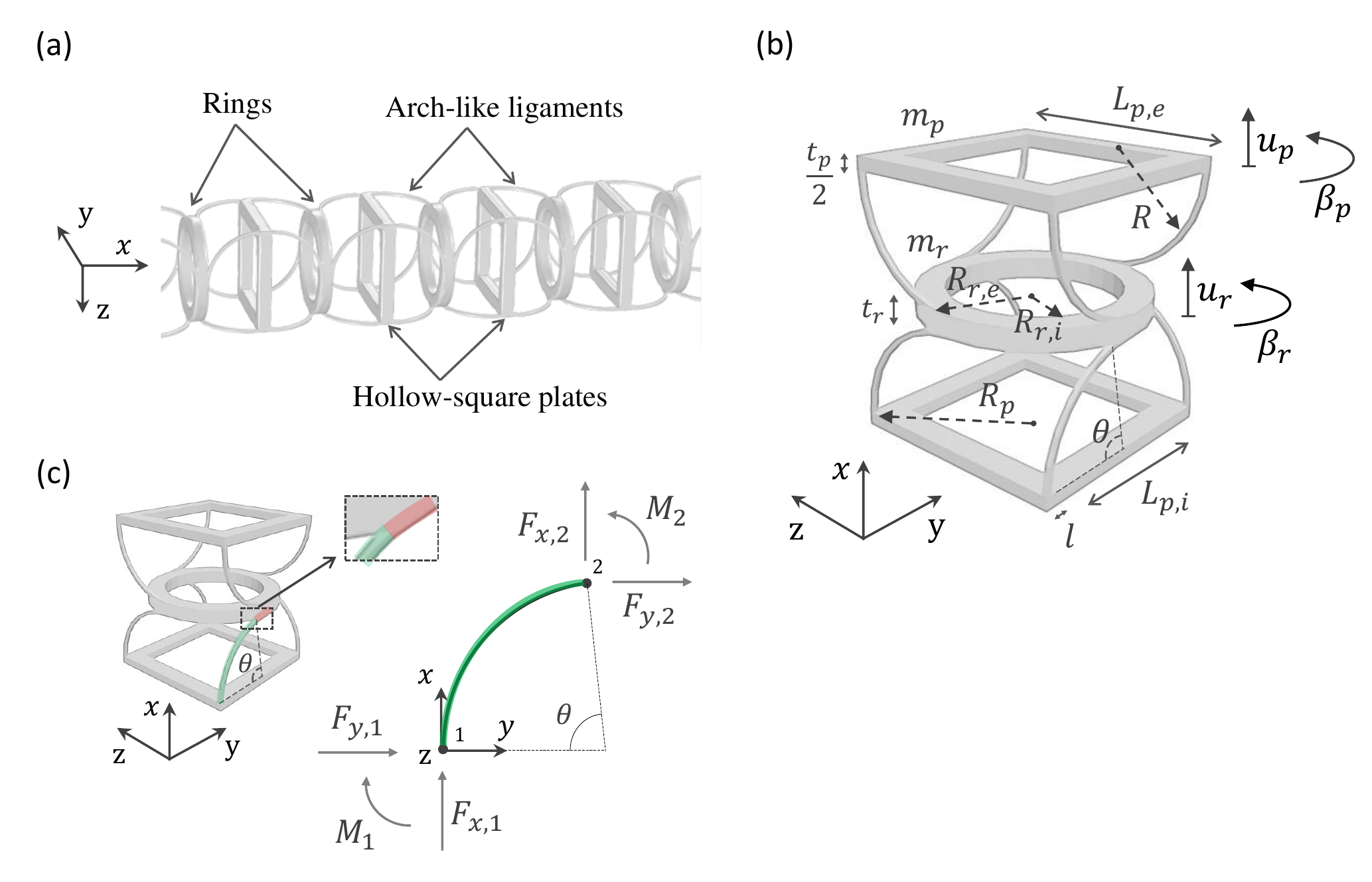}
\caption{\label{fig:model} | \textbf{Lattice building block.} (a) Inertial amplified metastructure with opposite chirality, comprised of an array of hollow-square plates connected to circular rings via arch-like ligaments. (b) Structure and geometric parameters of the building block. (c) Schematic of the analytical model adopted for the arch-like ligament. }
\end{figure}

Fig. \ref{fig:model}(a) shows the schematic of the inertial amplified 1D lattice with opposite chirality. Its building block consists of a hollow-square plate of mass $m_p$ connected to a circular ring of mass $m_r$ through four arch-like ligaments of radius $R$ (see Fig. \ref{fig:model}(b)). The curved shape of the ligaments enables the coupling between the axial and rotational motions of the mass elements. In particular, a relative motion along the $x$-direction triggers a relative rotation around the $x$-axis. 

We denote with $L_{p,e}$ and $L_{p,i}$ the external and internal side of the plate, respectively, $t_p$ designates the plate thickness, $R_{r,e}$ and $R_{r,i}$ the external and internal radius of the ring, $t_r$ the ring thickness, and, finally, $l$ the cross section dimension of the ligament. The external radius of the ring $R_{r,e}$ and the radius of curvature of the arch-like ligament $R$ are geometrically related as $R_{r,e}=R$. The geometric dimensions of the lattice unit cell, normalized with respect to the lattice constant $a$, are reported in Table \ref{t:properties}. The employed material is assumed to be stainless steel (Young’s modulus $E_s=210$ GPa, Poisson’s ratio $\nu_s=0.3$, and mass density $\rho_s=7850$ Kg/m$^3$), which is commonly employed to fabricate the coil-spring isolation systems for the attenuation of low-frequency mechanical vibrations \citep{Farrat}.

\begin{table}[H]
\centering
\begin{tabular}{c|ccccccc}
Quantity & $R_{r,e}$ & $R_{r,i}$ & $L_{p,e}$ & $L_{p,i}$ & $t_r$ & $t_p$ & $l$ \\
\hline
Dimensions & 0.4255$a$ & 0.3191$a$ & 0.8511$a$ & 0.6383$a$ & 0.0851$a$ & 0.1064$a$ & 0.0213$a$
\end{tabular}
\caption{Geometric dimensions of the unit cell components described in Fig. \ref{fig:model}(b). \label{t:properties}}
\end{table}

\section{Dispersion analysis of the infinite structure}
\subsection{Lumped parameter model}
To analytically calculate the dispersion diagram of the infinite structure, we consider a lumped parameter model. The hollow-square plate and the ring contribute to the mass and the moment of inertia of the system, while the arch-like ligaments are primarily responsible for the stiffness and elastic response. Under axial and/or torsional loading, both mass elements maintain an orientation parallel to the $y-z$ plane, implying that two DOFs per mass element, the rotational angle $\beta$ and the axial displacement $u$, are to be considered. The links are modelled according to the Euler-Bernoulli theory, assuming negligible shear deformation and curvature effects, since the assumed radius of curvature $R$ is large compared to the dimension of the beam cross-section $l$ \citep{Nambudiripad}. We further assume torsional effects to be negligible. In order to implement the Bloch theory, which allows us to derive the dispersion relation for the infinite lattice, the equilibrium of the $i^{th}$ block is examined. The equations of motion for the $i^{th}$ plate and the $i^{th}$ ring can be expressed as:
\begin{equation}
   m_p \ddot{u}_{p,i} = F(u_{r,i}-u_{p,i},\beta_{r,i}-\beta_{p,i}) - F(u_{p,i}-u_{r,i-1},\beta_{p,i}-\beta_{r,i-1})
  \label{eq:eq1}
\end{equation}
\begin{equation}
   I_p \ddot{\beta}_{p,i} = M_t(u_{r,i}-u_{p,i},\beta_{r,i}-\beta_{p,i}) - M_t(u_{p,i}-u_{r,i-1},\beta_{p,i}-\beta_{r,i-1})
  \label{eq:eq2}
\end{equation}
\begin{equation}
   m_r \ddot{u}_{r,i} = F(u_{p,i+1}-u_{r,i},\beta_{p,i+1}-\beta_{r,i}) - F(u_{r,i}-u_{p,i},\beta_{r,i}-\beta_{p,i})
  \label{eq:eq3}
\end{equation}
\begin{equation}
   I_r \ddot{\beta}_{r,i} = M_t(u_{p,i+1}-u_{r,i},\beta_{p,i+1}-\beta_{r,i}) - M_t(u_{r,i}-u_{p,i},\beta_{r,i}-\beta_{p,i})  
  \label{eq:eq4}
\end{equation}
where $u_{p,i}$, $\beta_{p,i}$, $u_{r,i}$, $\beta_{r,i}$ represent the axial and rotational displacements of the plate and the ring, respectively, $I_p$ and $I_r$ are the related moments of inertia, while $F$ and $M_t$ describe the axial and torsional elastic forces of the system provided by the arch-like ligaments.\\
The stiffness of the ligaments is obtained by applying Castigliano’s second theorem on the strain energy formulation \citep{Yamada}. The ligament is depicted in Fig. \ref{fig:model}(c). Its stiffness matrix, relating vertical ($F_{x,1}$, $F_{x,2}$) and horizontal ($F_{y,1}$, $F_{y,2}$) forces with the displacements ($u_{x,1}$, $u_{y,1}$, $u_{x,2}$, $u_{y,2}$), reads:
\begin{equation}
    \left[ {\begin{array}{c}
   F_{x1}\\
   F_{y1}\\
   F_{x2}\\
   F_{y2}\\
  \end{array} } \right] = D
   \left[ {\begin{array}{cccc}
   k_1 & -k_2 & -k_1 & k_2\\
   -k_2 & k_3 & k_2 & -k_3\\
   -k_1 & k_2 & k_1 & -k_2\\
   k_2 & -k_3 & -k_2 & k_3\\
  \end{array} } \right]
  \left[ {\begin{array}{c}
   u_{x1}\\
   u_{y1}\\
   u_{x2}\\
   u_{y2}\\
  \end{array} } \right] 
 \label{eq:system}
\end{equation}
where
\begin{gather}
  k_1= \varphi^2-\varphi(\cos \varphi ^2 - \sin \varphi ^2) \sin\varphi \cos\varphi - 2 \frac{\sin \varphi ^4}{(1+\xi)}  \nonumber
\end{gather}
\begin{gather}
  k_2=2\varphi \sin \varphi ^2 \cos \varphi ^2 - 2 \frac{\sin \varphi ^3 \cos \varphi}{(1+\xi)}  \nonumber\\
  k_3=\varphi^2+\varphi(\cos \varphi ^2 - \sin \varphi ^2) \sin\varphi \cos\varphi - 2 \frac{\sin \varphi ^2 \cos \varphi ^2}{(1+\xi)}  \nonumber\\
  D = \frac{E_sJ/R^3}{(1+\xi) \varphi (\varphi^2 - \sin \varphi ^2 \cos \varphi ^2) -2 (\varphi - \sin \varphi  \cos \varphi ) \sin \alpha ^2 }, \xi=\frac{J}{AR^2}, \varphi=\frac{\theta}{2} \nonumber
\end{gather}
The parameter $\xi$ corresponds to the ratio between the flexural and the axial stiffness, $\theta$ is the opening angle of the curved arch-like ligament, whereas $A$ and $J$ are the area and the moment of inertia of the cross-section, respectively. Assuming small displacements, the horizontal displacements at the ligaments ends are related to the rotational angle of the plate ($\beta_p$) and the ring ($\beta_r$) via: $u_{x,1}=(\sqrt2/2)\beta_p R_p$ and $u_{x,2}=\beta_r R_{r,e}$. Here, $R_p$ is the distance between the center of the square plate and its vertex, where the arch-like ligament is joined. The vertical displacements of the curved beam nodes, instead, coincide with the axial motion of the mass elements.\\
By substituting the elastic forces of Eq. \ref{eq:system} in Eqs. \ref{eq:eq1}, \ref{eq:eq2}, \ref{eq:eq3}, \ref{eq:eq4}, and by virtue of the Bloch boundary conditions \citep{Brillouin}, the equations of motion can be rewritten as:
\begin{equation}
   m_p \ddot{u}_{p,i} + 8D k_1 u_{p,i} - 4D k_1(1+\textrm{e}^{-\textrm{i}\gamma a}) u_{r,i} + 4D k_2 R_{r,e}(1-\textrm{e}^{-\textrm{i}\gamma a}) \beta_{r,i} = 0
  \label{eq:eq1m}
\end{equation}
\begin{equation}
   I_p \ddot{\beta}_{p,i} + 4D k_3 {R_p}^2 \beta_{p,i} + 2\sqrt{2}D k_2 R_p (1-\textrm{e}^{-i\gamma a}) u_{r,i} - 2\sqrt{2}D k_3 R_p R_{r,e} (1+\textrm{e}^{-\textrm{i}\gamma a}) \beta_{r,i} = 0
  \label{eq:eq2m}
\end{equation}
\begin{equation}
   m_r \ddot{u}_{r,i} - 4D k_1(1+\textrm{e}^{\textrm{i}\gamma a}) u_{p,i} + 2\sqrt{2}D k_2 R_p (1-\textrm{e}^{\textrm{i}\gamma a}) \beta_{p,i} + 8D k_1 u_{r,i} = 0
  \label{eq:eq3m}
\end{equation}
\begin{equation}
   I_r \ddot{\beta}_{r,i} + 4D k_2 R_{r,e}(1-\textrm{e}^{\textrm{i}\gamma a}) u_{p,i} - 2\sqrt{2}D k_3 R_p R_{r,e} (1+\textrm{e}^{\textrm{i}\gamma a}) \beta_{p,i} + 8D k_3 {R_{r,e}}^2 \beta_{r,i} = 0
  \label{eq:eq4m}
\end{equation}
where $\gamma$ designates the wavenumber, $\omega$ is the circular frequency, and $i$ denotes the imaginary unit.

The inertial amplification becomes evident after some algebra. We show how, after few substitutions, an additional inertia term arises in the first equation of motion (Eq. \ref{eq:eq1m}), relative to the translational motion of the plate.\\
From Eq. \ref{eq:eq4m}, we derive:
\begin{equation}
   \beta_{r,i} =  -\frac{I_r}{8D k_3 {R_{r,e}}^2} \ddot{\beta}_{r,i} - \frac{k_2 (1-\textrm{e}^{\textrm{i}\gamma a})}{2 k_3 R_{r,e}} u_{p,i} + \frac{\sqrt{2} R_p (1+\textrm{e}^{\textrm{i}\gamma a})}{4 R_{r,e}} \beta_{p,i}
  \label{eq:eq5}
\end{equation}
Via substitution of this result into Eq. \ref{eq:eq2m}, we obtain:
\begin{equation}
\begin{split}
    & \beta_{p,i} =  \frac{1}{\chi_1} \Big[ I_p \ddot{\beta}_{p,i} + \frac{\sqrt{2} R_p I_r (1+\textrm{e}^{-\textrm{i}\gamma a})}{4R_{r,e}} \ddot{\beta}_{r,i} \\[1.2ex]
    & + 2\sqrt{2}D k_2 R_p (1-\textrm{e}^{-\textrm{i}\gamma a}) u_{r,i} + \sqrt{2}D k_2 R_p (1+\textrm{e}^{-\textrm{i}\gamma a})(1-\textrm{e}^{\textrm{i}\gamma a}) u_{p,i} \Big]
\end{split}
  \label{eq:eq6}
\end{equation}
where $\chi_1$ reads:
\begin{align}
    \chi_1 = D k_3 {R_p}^2 \left[(1+\textrm{e}^{-\textrm{i}\gamma a})(1+\textrm{e}^{\textrm{i}\gamma a})-4 \right] 
\end{align}
Finally, by substituting Eqs. \ref{eq:eq5}, \ref{eq:eq6} into Eq. \ref{eq:eq1m}, we obtain the following equilibrium equation, which describes the axial motion of the plate:
\begin{equation}
    \begin{split}
        & m_p \ddot{u}_{p,i} + \frac{\chi_2}{\chi_1}I_p \ddot{\beta}_{p,i} + \Big[ \frac{\sqrt{2} R_p I_r (1+\textrm{e}^{-\textrm{i}\gamma a})}{4R_{r,e}} \frac{\chi_2}{\chi_1} - \frac{k_2 I_r (1-\textrm{e}^{-\textrm{i}\gamma a})}{2 k_3 R_{r,e}} \Big] \ddot{\beta}_{r,i} \\[1.2ex]
        & + 8D k_1 u_{p,i} - 4D k_1(1+\textrm{e}^{-\textrm{i}\gamma a}) u_{r,i} - \frac{2D {k_2}^2 (1-\textrm{e}^{-\textrm{i}\gamma a})(1-\textrm{e}^{\textrm{i}\gamma a})}{k_3} u_{p,i} \\[1.2ex]
        & - \frac{\chi_2}{\chi_1} \sqrt{2}D R_p k_2 \Big[2 (1-\textrm{e}^{-\textrm{i}\gamma a}) u_{r,i} + (1+\textrm{e}^{-\textrm{i}\gamma a})(1-\textrm{e}^{\textrm{i}\gamma a}) u_{p,i} \Big] = 0
    \end{split}
    \label{eq:eq7}
\end{equation}
where $\chi_2$ reads:
\begin{align}
    \chi_2 =  \sqrt{2}D k_2 R_p (1-\textrm{e}^{-\textrm{i}\gamma a})(1+\textrm{e}^{\textrm{i}\gamma a}) = 2\sqrt{2}D k_2 R_p \textrm{i} sin(\gamma a)
\end{align}
After simplifying Eq. \ref{eq:eq7}, we obtain:
\begin{equation}
\begin{split}
    & m_p \ddot{u}_{p,i} + \frac{\sqrt{2} I_p k_2 (\textrm{e}^{\textrm{i}\gamma a}+1)}{R_p k_3 (\textrm{e}^{\textrm{i}\gamma a}-1)} \ddot{\beta}_{p,i} + \frac{2 I_r k_2}{R_{r,e} k_3 (\textrm{e}^{\textrm{i}\gamma a}-1)} \ddot{\beta}_{r,i} \\[1.2ex]
    & - \frac{4D (k_1 k_3 - k_2^2) (1+\textrm{e}^{-\textrm{i}\gamma a})}{k_3} u_{r,i} +  \frac{8D (k_1 k_3 - k_2^2) }{k_3} u_{p,i}=0
\end{split}
  \label{eq:eq8}
\end{equation}
Due to the axial-torsional motion coupling, two additional inertia terms appear in Eq. \ref{eq:eq8}. However, only the term relative to the rotational acceleration of the ring ($\ddot{\beta}_{r,i}$) features a non-vanishing real part and, thus, increases the effective inertia of the system.

\begin{figure}[H]
\centering
\includegraphics[width=1\textwidth]
{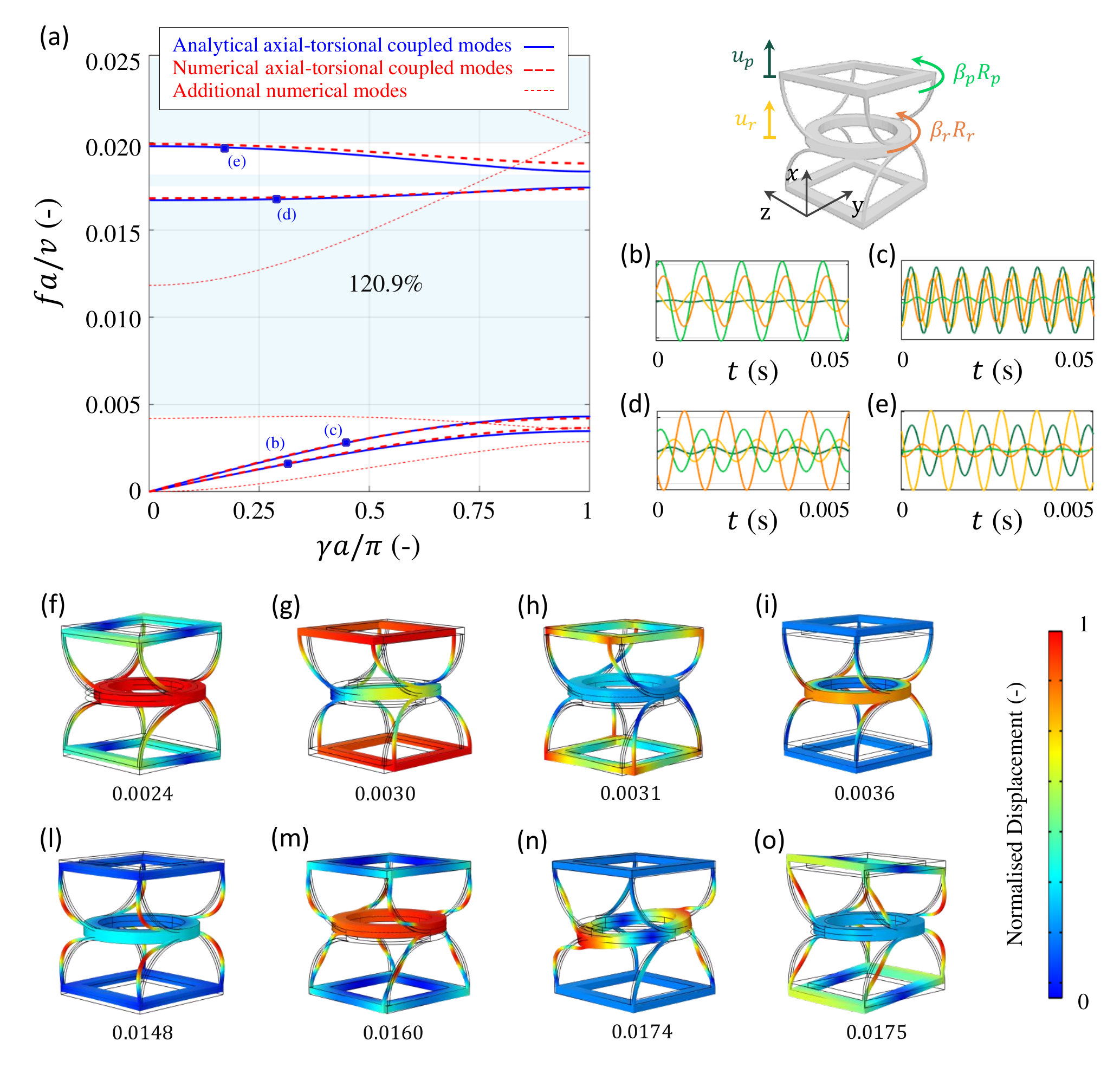}
\caption{\label{fig:disp} | \textbf{Dispersion analysis.} (a) Analytical (blue lines) and numerical (red dashed lines) band diagram of the lattice. (b)-(e) Four axial-torsional coupled mode shapes obtained from the analytical model. (f)-(o) Mode shapes obtained from the numerical model developed in COMSOL Multiphysics, which include the four axial-torsional coupled modes (h)-(i)-(l)-(m), corresponding to the analytical mode shapes (b)-(c)-(d)-(e).}
\end{figure}

Next, we derive the band diagram of the structure. By imposing a Bloch wave solution of the form $u_i= \bar u_i \textrm{e}^{\textrm{i}\gamma a-\textrm{i}\omega t}$, $\beta_i= \bar \beta_i \textrm{e}^{\textrm{i}\gamma a-\textrm{i}\omega t}$, the equations of motion (Eqs. \ref{eq:eq1m},\ref{eq:eq2m},\ref{eq:eq3m},\ref{eq:eq4m}) are rewritten in matrix form as:
\begin{equation}
\begin{split}
 \omega^2 & \left[ {\begin{array}{cccc}
   m_p & 0 & 0 & 0\\
   0 & I_p & 0 & 0\\
   0 & 0 & m_r & 0\\
   0 & 0 & 0 & I_r\\
  \end{array} } \right] 
  \left[ {\begin{array}{c}
   \bar u_p\\
   \bar \beta_p\\
   \bar u_r\\
   \bar \beta_r\\
  \end{array} } \right] =
  4D \left[\begin{matrix}
 2k_1 & 0\\
 0 & k_3 {R_p}^2 & \cdots\\
 -k_1(1+\textrm{e}^{\textrm{i}k a}) & (\sqrt{2}/2) k_2 R_p (1-\textrm{e}^{\textrm{i}k a}) & \cdots\\
 k_2 R_{r,e}(1-\textrm{e}^{\textrm{i}k a}) & -(\sqrt{2}/2) k_3 R_p R_{r,e} (1+\textrm{e}^{\textrm{i}k a})
\end{matrix}\right. \\[2ex]
& \left.\begin{matrix}
& -k_1(1+\textrm{e}^{-\textrm{i}k a}) & k_2 R_{r,e}(1-\textrm{e}^{-\textrm{i}k a})\\
\cdots& (\sqrt{2}/2) k_2 R_p (1-\textrm{e}^{-\textrm{i}k a}) & -(\sqrt{2}/2) k_3 R_p R_{r,e} (1+\textrm{e}^{-\textrm{i}k a})\\
\cdots& 2k_1 & 0\\
& 0 & 2k_3 {R_{r,e}}^2
\end{matrix}\right]
  \left[ {\begin{array}{c}
   \bar u_p\\
   \bar \beta_p\\
   \bar u_r\\
   \bar \beta_r\\
  \end{array} } \right] 
  \end{split} \label{eq:system_fin}
\end{equation}
In addition to increasing the inertial forces, the axial-torsional motion coupling generates four additional terms on the secondary diagonal of the wavenumber dependent stiffness matrix: $K(\gamma)_{14}$, $K(\gamma)_{23}$, $K(\gamma)_{32}$, $K(\gamma)_{41}$. Since $k_2$ is negative for $0<\theta<\pi/2$, these terms further reduce the elastic forces, thus lowering the resonant frequency of the system. The dispersion equation is derived from the solution of the eigenvalue problem: $\det (\mathbf{K}(\gamma)- \omega^2 \mathbf{M})=0$, considering the wavenumber $\gamma$ ranging from $0$ to $\pi/a$. Figure \ref{fig:disp}(a) shows the analytical band diagram (blue lines) of the structure in non-dimensional units. The non-dimensional frequency is defined as the ratio between the product of the frequency $f$ and the periodicity constant $a$ and the sound velocity of the material $v =\sqrt{E_s/\rho_s}$ \citep{Alessandro2}. Four axial-torsional coupled modes are visible. The lowest one essentially comprises a combination of rotary motion of the plate and axial-rotary motion of the disk (Fig. \ref{fig:disp}(b)). In the second lowest branch, the axial motion of the plate and the axial-rotary motion of the disk prevail (Fig. \ref{fig:disp}(c)). The third mode, mainly involves a rotation of the mass elements (Fig. \ref{fig:disp}(d)), while in the fourth mode both the plate and the disk mostly move axially (Fig. \ref{fig:disp}(e)). A wide bandgap, corresponding to a 120.9\% normalized width (defined as the ratio between the gap width and the mid-gap frequency $\Delta f/f_m$), arises between the second and the third mode. A second bandgap, less wide than the first (5.2\% normalized gap width), occurs between the third and fourth mode.

\subsection{Finite element model}
In order to now validate the analytical calculations, we construct a 3D finite element model of the lattice building block using COMSOL Multiphysics$^{\tiny{\raisebox{.5pt}{\textcircled{\scalebox{.8}{\raisebox{-0.4pt}{R}}}}}}$. We impose Bloch boundary conditions on the sides of the unit cell and perform an eigenfrequency analysis by sweeping the wavenumber in the range of 0 - $\gamma_\textrm{{max}}$ (with $\gamma_\textrm{{max}}=\pi/a$) to obtain the dispersion relation. Figure \ref{fig:disp}(a) shows the numerical band structure of the lattice (red dashed lines). Beyond the four axial-torsional coupled modes occurring under axial and/or torsional loading (see the bold dashed red lines in Fig. \ref{fig:disp}(a) and the mode shapes plotted in Figs. \ref{fig:disp}(h)-(i)-(l)-(m)), which are further captured by the analytical model, additional propagating modes are detected in the band diagram (dotted red lines in Fig. \ref{fig:disp}(a)).
These latter modes describe a translation of the plate along the $y$ or $z$ axes (Fig. \ref{fig:disp}(g)) and a rotation of both mass elements out of the $y-z$ plane (Figs. \ref{fig:disp}(f)-(n)-(o)). The slight mismatch between the four numerical and analytical axial-torsional coupled modes is attributed to the simplifying assumption of a lumped parameter model. In particular, in the numerical model, the ring mass constrains a small region of the ligament (red region in the link, shown in Fig. \ref{fig:model}(c)), thus slightly modifying the stiffness of the system. In the analytical model, this effect is accounted for by assuming the opening angle of the arch-like ligament $\theta$ smaller than $\pi/2$ (the opening angle returned by the numerical model is $\theta = 3\pi/7$).

\section{Bandgap mechanism}
\begin{figure}[H]
\centering
\includegraphics[width=1\textwidth]
{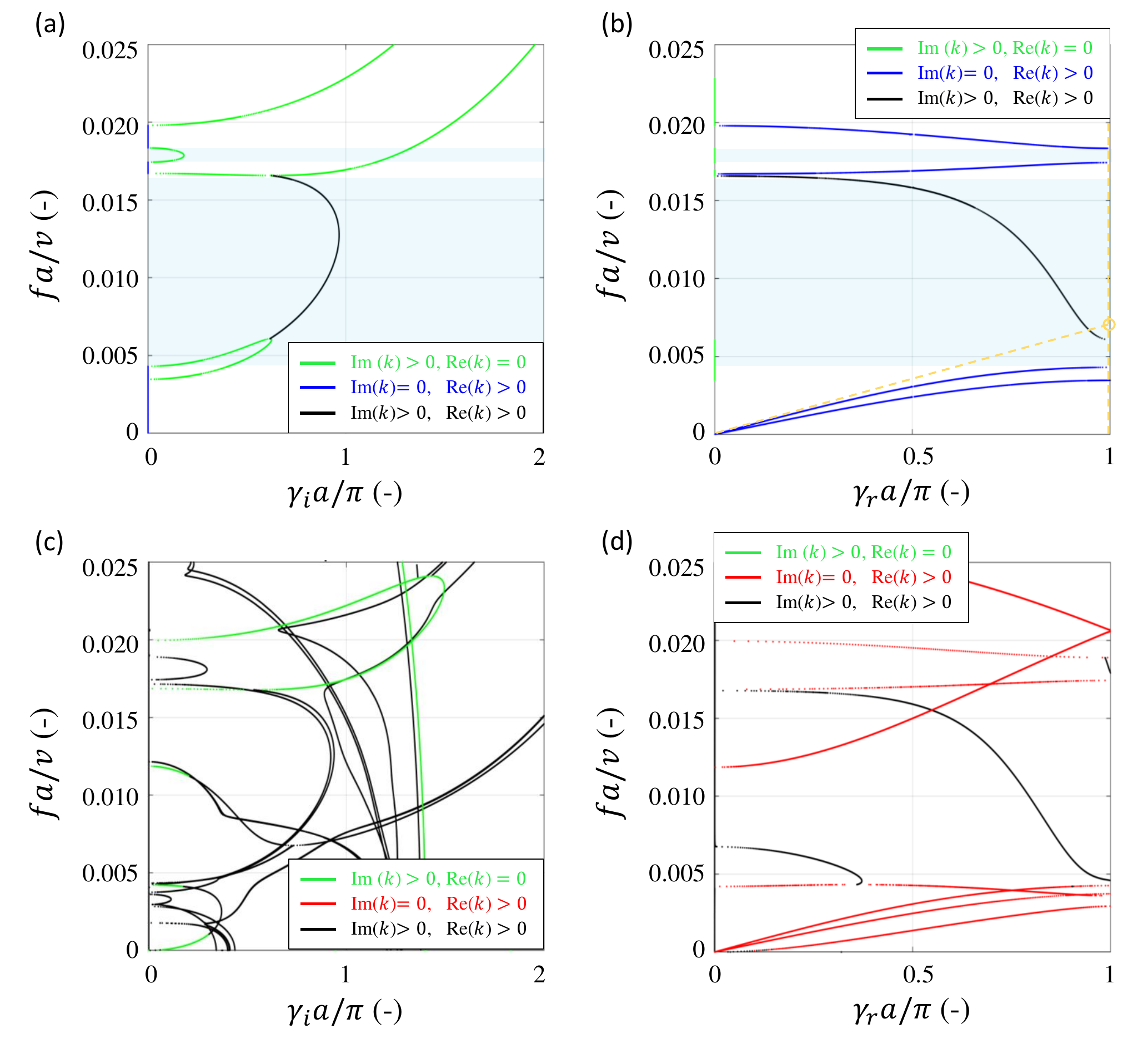}
\caption{\label{fig:disp_complex} | \textbf{Complex dispersion analysis.} Analytical (a)-(b) and numerical (c)-(d) dispersion curves for the inertial amplified structure in terms of the real and imaginary wavenumber components.  }
\end{figure}

To gain a deeper insight into the origin of the bandgap formation mechanism, we calculate the imaginary band structure. The calculation of the imaginary band structure requires a $\gamma(\omega)$ approach for the solution of the Bloch eigenvalue problem, where complex wavenumbers are sought for given real frequencies \citep{Laude}. The equilibrium equation $\big[\mathbf{K}(\gamma)- \omega^2 \mathbf{M}\big]\: \mathrm{u}=0$ can be rearranged as a quadratic eigenvalue problem. In particular, by denoting $\textrm{e}^{\textrm{i}\gamma a}=\mathrm{v}$ and $\textrm{e}^{-\textrm{i}\gamma a}=1/\mathrm{v}$, the equation is transformed to a characteristic polynomial in terms of variable $\mathrm{v}$, with coefficients that are functions of $\omega$, $\mathrm{M}$, and $\mathrm{K}$:
\begin{equation}
    [\mathrm{v}^2 \mathbf{A} + \mathrm{v} \mathbf{B} + \mathbf{A}^\mathrm{T}] \: \mathbf{u}=0
 \label{eq:system_imag}
\end{equation}
where
\begin{gather*}
   \mathbf{A} = D
   \left[ {\begin{array}{cccc}
   0 & 0 & 0 & 0\\
   0 & 0 & 0 & 0\\
   -k_1 & -(\sqrt{2}/2) k_2 R_p & 0 & 0\\
   -k_2 R_{r,e} & -(\sqrt{2}/2) k_3 R_p R_{r,e} & 0 & 0\\
  \end{array} } \right] \\
  \mathbf{B} = - \frac{\omega^2}{4} \mathbf{M} + D 
   \left[ {\begin{array}{cccc}
   2k_1 & 0 & -k_1 & k_2 R_{r,e}\\
   0 & k_3 R_p^2 & (\sqrt{2}/2) k_2 R_p & -(\sqrt{2}/2) k_3 R_p R_{r,e}\\
   -k_1 & (\sqrt{2}/2) k_2 R_p & 2k_1 & 0\\
   k_2 R_{r,e} & -(\sqrt{2}/2) k_3 R_p R_{r,e} & 0 & 2 k_3 R_{r,e}^2\\
  \end{array} } \right] 
\end{gather*}
Once the roots of the eigenvalue problem in Eq. (4) are computed, the real part $\gamma_r$ and the imaginary part $\gamma_i$ of the complex wavenumbers can be obtained as:
\begin{equation}
    \gamma_r = \frac{1}{a} \mathrm{arctan \frac{Im(v)}{Re(v)}}  \; \; \; \; \; \; \; \; \gamma_i = - \mathrm{\frac{ln \left|v\right|}{a}}
 \label{eq:imag}
\end{equation}
In contrast to the traditional $\omega(\gamma)$ problem, which assumes that $\gamma$ is real, the wavenumber vector is here considered as complex $\gamma=\gamma_{r}+\textrm{i}\gamma_{i}$. Figure \ref{fig:disp_complex} shows the dispersion curves in terms of imaginary (a) and real (b) wavenumber components, in non-dimensional units. While the real part of the wave vector is restricted to the first Brillouin zone, the imaginary part is unbounded. The wavenumbers are classified into three groups: (i) wavenumbers with a real part that lies inside the Brillouin zone, while the imaginary part equals zero ($\gamma_r a/\pi<1,\: \gamma_i=0$) are indicated via blue lines; (ii) pure imaginary wavenumbers ($\gamma_r=0, \: \gamma_i>0$) are indicated via green lines; (iii) complex wavenumbers characterized by both real and imaginary non-vanishing parts ($\gamma_r>0, \: \gamma_i>0$) are shown in black lines. As expected, the 

\noindent
$f-k$ pairs characterized by a pure real wavenumber (blue lines in Fig. \ref{fig:disp_complex}(b)) perfectly overlap with the real band diagram illustrated in Fig. \ref{fig:disp}(a). Interestingly, within the first bandgap identified by the theoretical model, a mode characterized by non-zero real and imaginary parts arises (black curve). This mode cannot emerge when solving the classical eigenvalue problem, where only pure real wave vectors are considered. However, since its imaginary part dominates over the real one, the propagation of this mode is expected to be strongly attenuated. The profile assumed by the pure imaginary modes (green curves) indicates that the bandgaps occur due to the Bragg scattering mechanism. Indeed, a typical Bragg gap is characterized by a continuous and smooth variation of the imaginary part of the wave vector \citep{Ao,Yuan}. Moreover, the Bragg mid-gap frequency, obtained by the intersection of the vertical line $\gamma_r a/\pi=1$ with the tangent to the second coupled axial-torsional mode, lies within the first bandgap (see the dashed yellow lines in Fig. \ref{fig:disp_complex}(b)). This confirms that the Bragg scattering controls the gap formation mechanism.

To validate the analytical findings, we numerically compute the complex band diagram by using the numerical, Finite Element (FE) - based $\gamma(\omega)$ approach described in \citep{Collet}. This method applies a Bloch operator transformation of the governing differential equation, constructs its weak form, and discretizes this over the unit cell domain using finite elements. The implementation is performed with COMSOL Multiphysics$^{\tiny{\raisebox{.5pt}{\textcircled{\scalebox{.8}{\raisebox{-0.4pt}{R}}}}}}$, once again adopting the numerical model, which was previously developed for the dispersion analysis. Figure \ref{fig:disp_complex} shows the dispersion curves numerically computed in terms of imaginary (c) and real (d) wavenumber components. Among the several modes, the numerical model returns also the imaginary and complex modes extracted from the analytical formulation (Figs. \ref{fig:disp_complex}(a)-(b)).

\section{Wave propagation through finite-sized structures}
\begin{figure}[H]
\centering
\includegraphics[width=1\textwidth]
{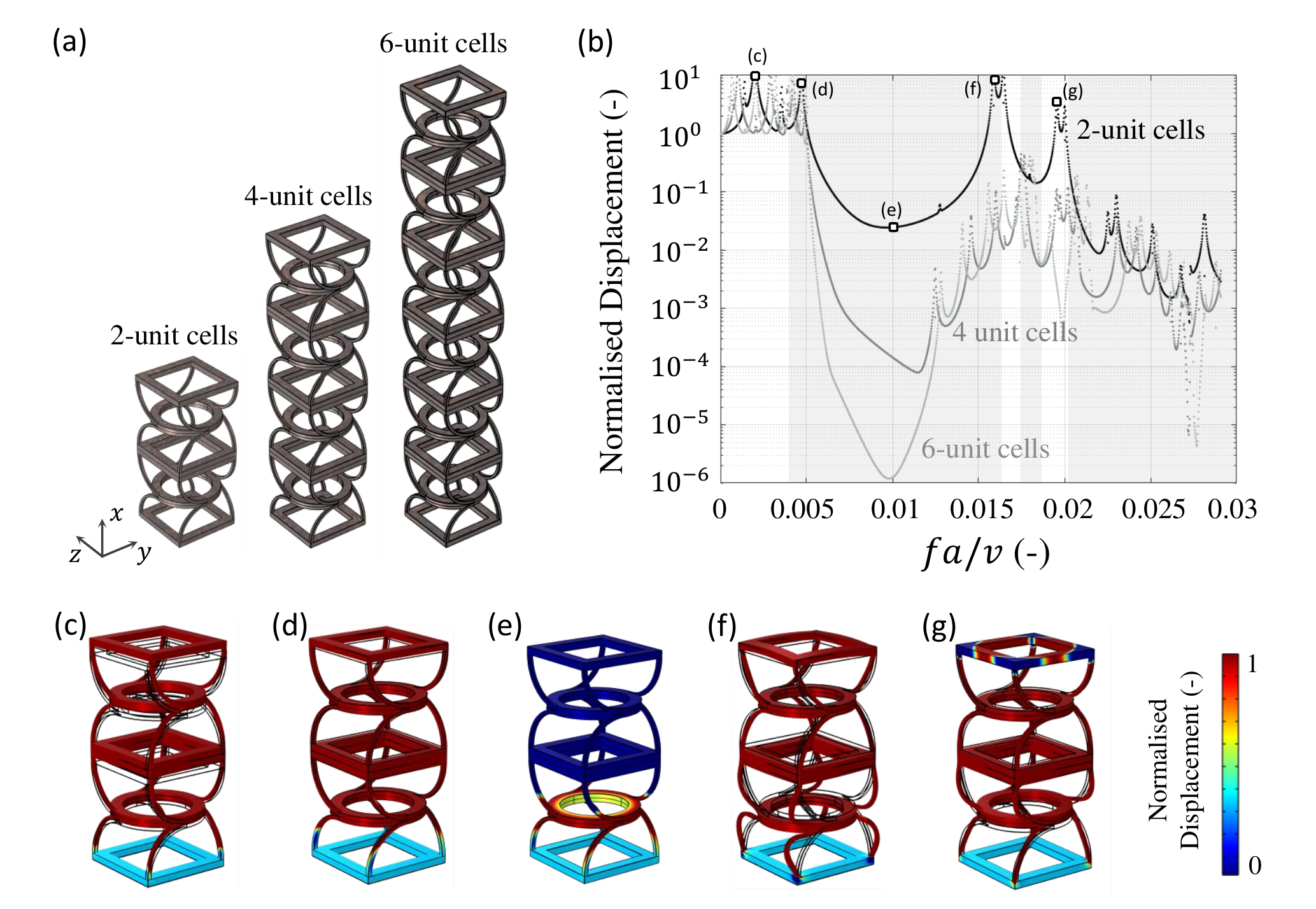}
\caption{\label{fig:freq} | \textbf{Wave transmission analysis.} (a) Model of the metastructure comprised of 2, 4, and 6 unit cells, respectively. (b) Normalized transmitted displacements $\sqrt{u_x^2+u_y^2+u_z^2}$ as a function of the frequency $f_\textrm{{nd}}$. The curves indicate the displacement values averaged on the free top end of the three increasing length metastructures with 2 (black lines), 4 (dark grey lines), and 6 (light grey lines) unit cells, respectively. Vibration mode at $f_\textrm{{nd}}=0.002$ (c), $f_\textrm{{nd}}=0.005$ (d), $f_\textrm{{nd}}=0.010$ (e), $f_\textrm{{nd}}=0.016$ (f), $f_\textrm{{nd}}=0.020$ (g), where $f_\textrm{{nd}}$ is the non-dimensional frequency. }
\end{figure}

In this section, we numerically investigate the linear elastic wave propagation in finite-sized structures with an increasing cell number. Figure \ref{fig:freq}(a) depicts the models of the three analysed structures, composed respectively of 2, 4, and, 6 unit cells. The geometric parameters of the building block of each structure are those employed in the dispersion analysis. We perform harmonic simulations with an imposed displacement applied to the base of the structures, along the $x$ axis. A material loss factor of $\xi=2 \times 10^{-3}$, typical of steel structures is considered. Figure \ref{fig:freq}(b) shows the amplitude of the total transmitted displacements $\sqrt{u_x^2+u_y^2+u_z^2}$ normalized with respect to the applied excitation $u_{x_0}$. 

\noindent
The black, dark gray and light gray curves correspond to the displacements computed on the free top end of the structure comprised of 2, 4, and 6 unit cells, respectively. The gray-shaded zones of Fig. \ref{fig:freq}(b) specify the respective bandgap regions, as identified in the dispersion study. We observe a good agreement between these zones and the frequency regions, where the wave transmission through the finite-sized structure is lower than 1. The wave attenuation intensifies as the number of unit cells forming the metastructure increases, particularly within the frequency region of the first bandgap. It is noteworthy that a metastructure with only two cells yields a reduction in the transmitted signal of almost two orders of magnitude (Fig. \ref{fig:freq}(e)). The peaks preceding the bandgaps in the transmission spectrum correspond to the axial-torsional resonant modes of the finite metastructure (see Figs. \ref{fig:freq}(c)-(d)-(f)-(g)). It is worth noting that, when 4 and 6 unit cells are considered, these peaks appear considerably attenuated. Indeed, these modes are not effectively activated as a result of their local nature. Consequently, the second and third bandgaps are combined and the attenuation properties of the structure are enhanced.

\section{Experimental validation}
\begin{figure}[H]
\centering
\includegraphics[width=1\textwidth]
{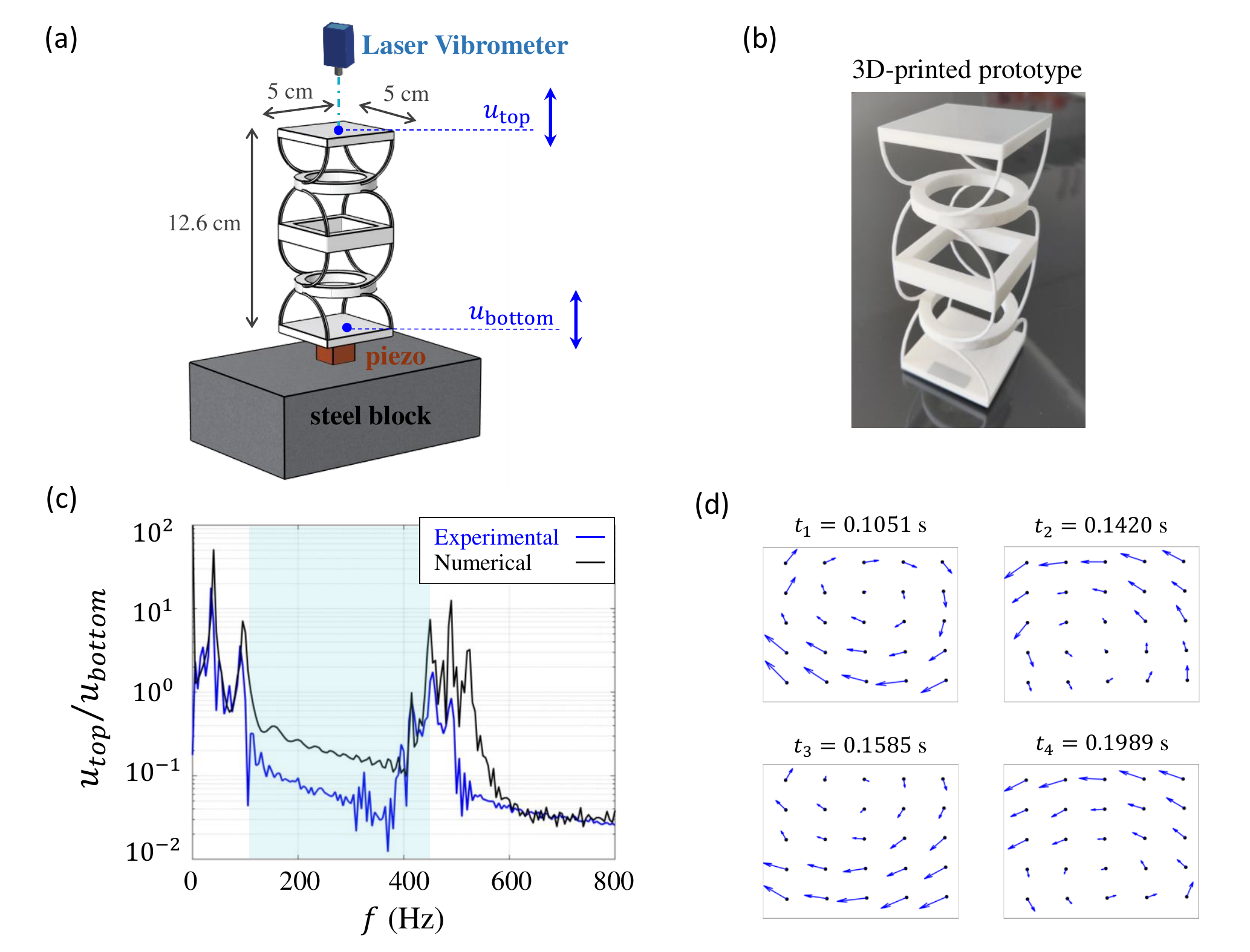}
\caption{\label{fig:setup} | \textbf{Experimental test.} (a) The experimental setup, involving the laser Doppler vibrometer, a three-axis piezoelectric actuator, and the inertial amplified metastructure. (b) Illustration of the 3D-printed prototype, made out of Nylon 12. (c) Experimental (blue lines) and numerical (black lines) vertical wave transmission spectra. The shaded light-blue area identifies the frequency range of the lowest bandgap. (d) Normalised in-plane displacement measured over a predefined grid of points on the top plate of the structure at four sequential time instants ($t_1=0.1051$ s, $t_2=0.1420$ s, $t_3=0.1585$ s, $t_4=0.1989$ s). The arrows indicate the direction of the rotation of the top plate. }
\end{figure}

To confirm the attenuation properties of the metastructure and experimentally validate its inertial amplified mechanism, we conduct an experimental test. We fabricate a small-scale prototype made of 2 unit cells, 12.6 cm high and 5 cm $\times$ 5 cm wide, in Nylon 12 ($E_n=1700$ MPa, $\nu_n=0.3$, $\rho_n=450$ Kg/m$^3$) by means of Selective Laser Sintering (see Fig. \ref{fig:setup}(b)). As shown in Fig. \ref{fig:freq}(b), the chosen number of cells allows for reducing the wave transmission at nearly two orders of magnitude, as well as limiting the dimensions of the finite-sized structure. As this is more convenient for 3D fabrication, and we here are dealing with a scaled-down proof of concept, we conduct the experimental test on a prototype made out of Nylon 12, a flexible and cheap material easy to print with commercial 3D printers. The geometric dimensions of the prototype are reported in Table \ref{t:proto}.

\begin{table}[H]
\centering
\begin{tabular}{c|ccccccc}
Quantity & $R_{r,e}$ & $R_{r,i}$ & $L_{p,e}$ & $L_{p,i}$ & $t_r$ & $t_p$ & $l$ \\
\hline
Dimensions (cm) & 2.5 & 1.875 & 5 & 3.75 & 0.6 & 1 & 0.12
\end{tabular}
\caption{Geometric dimensions of the 3D-printed prototype. \label{t:proto}}
\end{table}

As illustrated in Fig. \ref{fig:setup}(a), we vertically excite the structure at the base using a three-axis piezoelectric actuator, which is driven by a PiezoDrive amplifier. To do this, we activate only the piezoelectric stack corresponding to the vertical $x$-axis. A frequency sweep lasting 0.2 s with constant amplitude from 50 to 800 Hz is employed as excitation signal. We use a Polytech 3D laser Doppler vibrometer to measure the displacements over a predefined grid of points on the top plate and at a point on the bottom plate edge (reference input displacement $u_\textrm{{bottom}}$). The measurements are carried out with small amplitude excitations to ensure that the material features a linear elastic behaviour. The experimental wave transmission spectrum, obtained as the ratio between the output and input spectral displacements ($u_\textrm{{top}}$/$u_\textrm{{bottom}}$), is represented as solid blue line in Fig. \ref{fig:setup}(c). The light-blue-shaded zone highlights the frequency region of the lowest-frequency bandgap calculated via numerical dispersion analysis, which extends from 100 to 450 Hz. The curve exhibits a wide signal reduction starting from 100 Hz, with a decrease in the amplitude of vibration at almost 2 orders of magnitude, thus matching the results of the numerical frequency-domain analyses (Fig. \ref{fig:freq}(b)). The bandgap region is preceded by a first passband, where two slightly amplified peaks are visible at 35 and 95 Hz. As shown in Fig. \ref{fig:setup}(d), the peaks are generated as a result of the activation of the axial-torsional coupled modes. In particular, the figure depicts the normalised in-plane displacements, measured over the grid of points on the top plate, at four sequential time instants ($t_1 = 0.1051$ s, $t_2 = 0.1420$ s, $t_3 = 0.1585$ s, $t_4 = 0.1989$ s). The arrows indicate the direction of the rotation of the top plate. Finally, it is worth noting that, unlike the first passband, the second one is not effectively activated and the two bandgaps almost merge. This is attributed to the viscous damping featured by the adopted Nylon 12 material.

The solid black line in Fig. \ref{fig:setup}(c) represents the numerical transmission spectrum calculated by performing a time-transient simulation in COMSOL Multiphysics$^{\tiny{\raisebox{.5pt}{\textcircled{\scalebox{.8}{\raisebox{-0.4pt}{R}}}}}}$. In the numerical simulation, we model the material as linear elastic and we impose a loss factor of $\nu=0.05$, typical of Nylon 12. Numerical and experimental transmission spectra are in good agreement, although the experimental setup exhibits a higher attenuation level both in the bandgaps regions and in the high frequency regime. This can be ascribed to the material damping, whose experimental value may differ from that assumed in the numerical simulation.

\section{Stress-constrained optimization study}
Inertial amplification mechanisms are often realized by means of thin flexure hinges \citep{Acar,Taniker2} and links \citep{Orta,Wang}. As a result, these flexible structures may feature high internal stress values that can compromise their functionality. For such configurations, which are meant to be used at possibly resonant state, structural safety by preventing plastic deformations and static failure is thus of essence.

In what follows, we determine the optimal design parameters of the steel metastructure, comprised of 2 unit cells, which lead in attaining the lowest-frequency bandgap, while ensuring that the structure undergoes limited stress. To this end, we perform a stress-constrained optimization study, where the objective function minimizes the bandgap opening frequency, while the constraints lead in restriction of both the dimensions of the unit cell as well as the maximum stress value that can occur within the structure. To analyse how the optimal bandgap opening frequency is affected upon variation of the maximum admissible stress value, we conduct 20 constrained optimization problems using a different maximum admissible stress value $\overline{\sigma}=\{\overline{\sigma_1}, \overline{\sigma_2},...,\overline{\sigma_j}, ..., \overline{\sigma_{20}}\}$ in the stress constraint.\\
The steps taken to perform the stress-constrained optimization study are:

\begin{enumerate}
  \item Derive the objective function $f_\textrm{{bg}}$ from the analytical solution of the eigenvalue problem (Eq. \ref{eq:system_fin}).
  \item Estimate the stress constraint function $\hat{\sigma}^{\tiny{\textrm{vM}}}_\textrm{{max}}$ by performing a multivariate polynomial regression on the von Mises stress dataset obtained from the numerical model.
  \item For each of the imposed maximum admissible stress values ($\overline{\sigma_j}$), solve the constrained optimization problem using the sequential quadratic programming (SQP) method.
\end{enumerate}

\subsection{The optimization model}
The $j^{th}$ optimization problem can be mathematically formulated as follows:
\setlength{\belowdisplayskip}{0pt} \setlength{\belowdisplayshortskip}{0pt}
\setlength{\abovedisplayskip}{0pt} \setlength{\abovedisplayshortskip}{0pt}
\begin{flalign}
 \mbox{Find: \space} \mathcal{P} = \Big\{ R_{r,e} , l , t_p , t_r \Big\} \subset \mathbb{R}^4&& \label{opt}
\end{flalign}
\begin{flalign}
 \mbox{\textit{Minimize}: \space} f_\textrm{{bg}}\big(\mathcal{P}\big)&&  \label{obj}
\end{flalign}
\begin{flalign}
 \mbox{\textit{Subject to}: \space} \hat{\sigma}^{\tiny{\textrm{vM}}}_\textrm{{max}}\big(\mathcal{P}\big)<\overline{\sigma_j}&& \label{str}
\end{flalign}
\begin{flalign}
 \mbox{\hspace*{12.7ex}} R^\textrm{{lb}}_{r,e} <R_{r,e}< R^\textrm{{ub}}_{r,e} && \label{constr1}
\end{flalign}
\begin{flalign}
 \mbox{\hspace*{12.7ex}} l^\textrm{{lb}} <l<l^\textrm{{ub}} && \label{constr2}
\end{flalign}
\begin{flalign}
 \mbox{\hspace*{12.7ex}} t^\textrm{{lb}}_{p} <t_p<t^\textrm{{ub}}_{p} && \label{constr3}
\end{flalign}
\begin{flalign}
 \mbox{\hspace*{12.7ex}} t^\textrm{{lb}}_{r} <t_r<t^\textrm{{ub}}_{r} && \label{constr4}
\end{flalign}
where $\mathcal{P}$ denotes the vector of design parameters to be optimized, namely the external radius of the ring $R_{r,e}$, the cross section dimension of the ligament $l$, and the thickness of both mass elements ($t_p$, $t_r$). $f_\textrm{{bg}}\big(\mathcal{P}\big) : \mathbb{R}^4 \rightarrow \mathbb{R}$ designates the function determining the opening frequency of the bandgap and $\hat{\sigma}^{\tiny{\textrm{vM}}}_\textrm{{max}} : \mathbb{R}^4 \rightarrow \mathbb{R}$ denotes the function defining the maximum von Mises stress value occurring in the structure. $\overline{\sigma_j}$ is the maximum permissible stress. The lower (lb) and upper (up) bounds of the variables are listed in Table \ref{t:opt}.

\begin{table}[H]
\centering
\begin{tabular}{c|cccc}
Variables & $R_{r,e}$ (cm) & $l$  (cm) & $t_r$  (cm) & $t_p$  (cm) \\
\hline
Lower bound (lb) & 6 & 0.2 & 1 & 1  \\
Upper bound (ub) & 10 & 0.5 & 5 & 5
\end{tabular}
\caption{Upper and lower bounds of the variables $R_{r,e}$, $l$, $t_r$, $t_p$. \label{t:opt}}
\end{table}

The function $f_\textrm{{bg}}$ defining the frequency of the second axial-torsional coupled mode, which determines the position of the bandgap lower bound, is employed as objective function (Eq. \ref{obj}) in the optimization study. This is derived from the analytical solution of the eigenvalue problem (Eq. \ref{eq:system_fin}) considering the wavenumber $\gamma=\pi/a$.

\subsection{Stress constraint}
\begin{figure}[H]
\centering
\includegraphics[width=1\textwidth]
{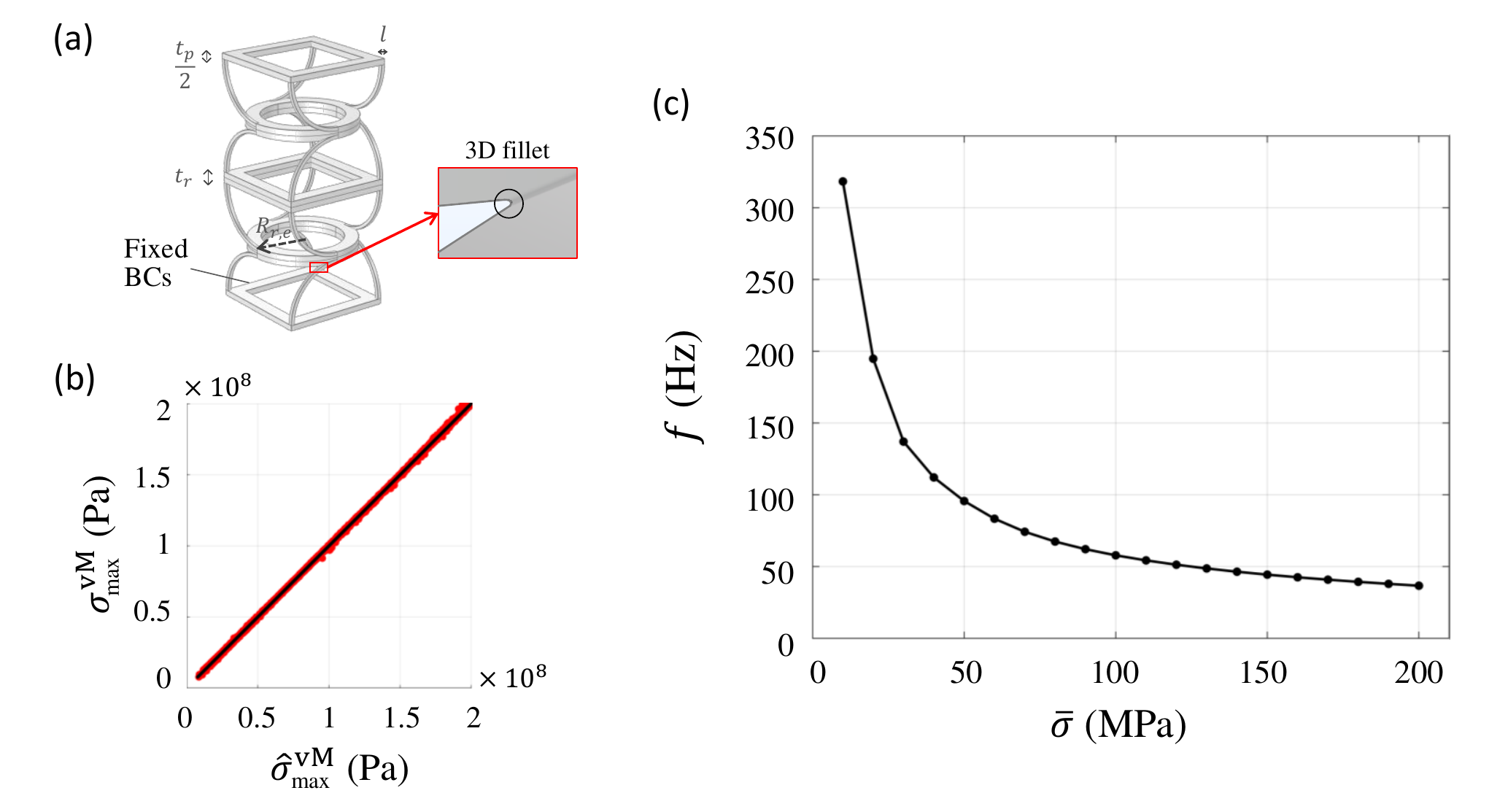}
\caption{\label{fig:opt} | \textbf{Optimization Problem.} (a) Model of the metastructure comprised of 2 unit cells for the linear static analysis. (b) Scatter plot showing how the predicted stress values ($\hat{\sigma}^{\tiny{\textrm{vM}}}_\textrm{{max}}$, on the $x$-axis) agree well with the observed ones ($\sigma^{\tiny{\textrm{vM}}}_\textrm{{max}}$, on the $y$-axis). (c) Bandgap opening frequency as a function of the maximum permissible stress value.}
\end{figure}

The stress inequality constraint (Eq. \ref{str}) limits the maximum value of von Mises stress that can occur in the structure under static conditions. The nonlinear function $\hat{\sigma}^{\tiny{\textrm{vM}}}_\textrm{{max}}\big(\mathcal{P}\big)$, which describes how the maximum internal stress value changes upon variation of the design parameters, is derived by performing a multivariate polynomial regression \citep{Cecen}. The independent variables are the design parameters, i.e., the external radius of the ring, the cross section dimension of the ligament, and the thickness of both mass elements contained in the vector $\mathcal{P}$. The dependent variable is the maximum von Mises stress value occurring in the structure. We generate the training set using the Latin Hypercube Sampling (LHS) method, which ensures that the adopted set of random numbers is representative of the real variability. In particular, we generate a near-random sample of the design parameter values using the LHS method. Next, we obtain the observed values of the dependent variable by numerically computing the maximum von Mises stress values resulting from each combination of the design parameter sample. In particular, for each parametric set combination that is generated by the LHS, we develop a finite element model using COMSOL Multiphysics$^{\tiny{\raisebox{.5pt}{\textcircled{\scalebox{.8}{\raisebox{-0.4pt}{R}}}}}}$ (see Fig. \ref{fig:opt}(a)). To avoid local singularities in the simulations, we add 3D fillets that round off the sharp edges, which form when the ligaments join the plates and the rings (see the inset of Fig. \ref{fig:opt}(a)). While the bottom end of the structure is fixed, the top is free to move. We use tetrahedral elements satisfying mesh convergence. For each parameter set, we perform a linear static analysis and we compute the internal stresses occurring in the structure. Finally, we extract the maximum von Mises stress values from the simulations and use them as observed values of the dependent variable in the multivariate polynomial regression to derive the approximate form of the analytical stress function $\hat{\sigma}^{\tiny{\textrm{vM}}}_\textrm{{max}}\big(\mathcal{P}\big)$.

We employ the least squares method to fit the polynomial regression model. The degree of the polynomial regression is chosen via leave-one-out cross-validation. Based on the cross-validation results, we select the 4$^{th}$ degree. To measure the goodness of the fit, we calculate the R-Squared and the mean absolute error (MAE) as R$^2=0.99$, MAE$=1.5\cdot 10^{-3}$. The scatter plot, depicted in Fig. \ref{fig:opt}(b), indicates the relationship between the predicted stress values ($\hat{\sigma}^{\tiny{\textrm{vM}}}_\textrm{{max}}$, on the $x$-axis) and the observed ones ($\sigma^{\tiny{\textrm{vM}}}_\textrm{{max}}$, on the $y$-axis). The red points located along the black diagonal line confirm the goodness of the fitted regression model.

We then use the fitted stress function $\hat{\sigma}^{\tiny{\textrm{vM}}}_\textrm{{max}}\big(\mathcal{P}\big)$ to build the stress constraint (Eq. \ref{str}) of the optimization problem that limits the maximum permissible stress. As mentioned at the beginning of the section, the latter assumes different values in each of the 20 optimization problems. The aims of these analyses is to investigate how the bandgap lower limit and the metastructure design parameters are affected upon its variation. $\overline{\sigma_j}$ assumes values ranging from 10 to 200 MPa. Depending on the specifications the isolation system has to meet and the possible external load, the internal stresses caused by the structure's self-weight must be limited to ensure that the metastructure operates in the elastic regime. The values assumed by $\overline{\sigma_j}$ are reported in Table \ref{t:opt_arch} as a percentage of the steel yielding stress $\sigma_Y=230$ MPa. By maintaining the internal stresses below 200 MPa, the structure undergoes limited elastic deformations (strains $\varepsilon<2$\textperthousand). Note that, by performing pre-stressed eigenfrequency analyses, we verify that the increasing internal stresses, due to the structure self-weight, do not affect the band diagram.

\subsection{Constrained optimization study}
We impose constraints (Eqs. \ref{constr1}, \ref{constr2}, \ref{constr3}, \ref{constr4}), with limit values described in Table \ref{t:opt}, over the design parameters to avoid either large-sized or extremely small structures.

We employ the sequential quadratic programming (SQP) method to solve each of the 20 optimization problems. From an initial set of geometric parameters $\mathcal{P}_0$, the method performs a series of iterative steps to minimize the objective function, while keeping the specified constraints satisfied. The convergence is reached and the iteration process stops when the step becomes smaller than the step relative tolerance.\\
The SQP method requires specification of an initial starting point $\mathcal{P}_0$ for the optimization process. In an attempt to avoid entrapment in a local minimum, we generate a near-random sample of initial points using the latin hypercube sampling method. We then find an optimal solution for each guess point. Next, we compare all the solutions one to another to identify the optimal geometric parameters.\\
The results are reported in Table \ref{t:opt_arch} and illustrated in Fig. \ref{fig:opt}(c). The bandgap opening frequency $f_\textrm{{bg}}$ decreases exponentially with increasing maximum permissible stress $\overline{\sigma}$. For $\overline{\sigma}=200$ MPa, the lowest achievable $f_{bg}$ is 36.7 Hz. This is obtained by designing a structure comprising 2 unit cells of dimension $a=t_p+t_r+2(R_{r,e}-l)=21.46$ cm.

Alongside specification of the relationship between the bandgap opening frequency and the maximum permissible stress value, it is worth examining the values assumed by the design parameters. The geometric variables that most affect $f_\textrm{{bg}}$ are the external radius of the ring $R_{r,e}$ and the cross section dimension of the ligament $l$. On the contrary, the thickness of the mass elements plays a minor role in the decrease of $f_\textrm{{bg}}$. Indeed, $t_r$ and $t_p$ always assume the lowest possible value, which is identified by the lower bound of the parameter interval ($1$ cm, see constraints \ref{constr3},\ref{constr4}). The different influence that the design parameters have on the bandgap opening frequency can be explained by observing how they affect the inertia of the system. While the rotational inertia of the system exhibits a quadratic growth as the external radius of the ring increases ($I_p,I_r\propto R_{r,e}^2$), the thickness of the mass elements contributes linearly to the overall mass of the structure ($m_r\propto t_r$; $m_p\propto t_p$). This causes the external radius of the ring to assume the highest possible value, as permitted by the margins imposed by the stress constraint.

\begin{table}[H]
\centering
\begin{tabular}{c|ccccc}
$\overline{\sigma}$ (MPa) & $f_\textrm{{bg}}$ (Hz)  & $R_{r,e}$ (cm) & $l$ (mm) & $t_p$ (cm) & $t_r$ (cm) \\
\hline
$10=$ $\sim 4 \tiny{\%}$ $\sigma_Y$ & 318.23 & 6.65 & 4.63 & 1.00 & 1.00 \\
$20=$ $\sim 9 \tiny{\%}$ $\sigma_Y$ & 194.78 & 8.28 & 4.80 & 1.00 & 1.00 \\
$30=$ $\sim 13 \tiny{\%}$ $\sigma_Y$ & 137.05 & 9.81 & 5.00 & 1.00 & 1.00 \\
$40=$ $\sim 17 \tiny{\%}$ $\sigma_Y$ & 112.11 & 10.00 & 4.64 & 1.00 & 1.00 \\
$50=$ $\sim 22 \tiny{\%}$ $\sigma_Y$ & 95.69 & 10.00 & 4.30 & 1.00 & 1.00 \\
$60=$ $\sim 26 \tiny{\%}$ $\sigma_Y$ & 83.36 & 10.00 & 4.02 & 1.00 & 1.00 \\
$70=$ $\sim 30 \tiny{\%}$ $\sigma_Y$ & 74.26 & 10.00 & 3.80 & 1.00 & 1.00 \\
$80=$ $\sim 35 \tiny{\%}$ $\sigma_Y$ & 67.45 & 10.00 & 3.62 & 1.00 & 1.00 \\
$90=$ $\sim 39 \tiny{\%}$ $\sigma_Y$ & 62.14 & 10.00 & 3.48 & 1.00 & 1.00 \\
$100=$ $\sim 43 \tiny{\%}$ $\sigma_Y$ & 57.87 & 10.00 & 3.36 & 1.00 & 1.00 \\
$110=$ $\sim 49 \tiny{\%}$ $\sigma_Y$ & 54.33 & 10.00 & 3.26 & 1.00 & 1.00 \\
$120=$ $\sim 52 \tiny{\%}$ $\sigma_Y$ & 51.33 & 10.00 & 3.17 & 1.00 & 1.00 \\
$130=$ $\sim 57 \tiny{\%}$ $\sigma_Y$ & 48.73 & 10.00 & 3.09 & 1.00 & 1.00 \\
$140=$ $\sim 61 \tiny{\%}$ $\sigma_Y$ & 46.44 & 10.00 & 3.02 & 1.00 & 1.00 \\
$150=$ $\sim 65 \tiny{\%}$ $\sigma_Y$ & 44.40 & 10.00 & 2.95 & 1.00 & 1.00 \\
$160=$ $\sim 70 \tiny{\%}$ $\sigma_Y$ & 42.57 & 10.00 & 2.89 & 1.00 & 1.00 \\
$170=$ $\sim 74 \tiny{\%}$ $\sigma_Y$ & 40.91 & 10.00 & 2.84 & 1.00 & 1.00 \\
$180=$ $\sim 78 \tiny{\%}$ $\sigma_Y$ & 39.39 & 10.00 & 2.78 & 1.00 & 1.00 \\
$190=$ $\sim 83 \tiny{\%}$ $\sigma_Y$ & 37.99 & 10.00 & 2.73 & 1.00 & 1.00 \\
$200=$ $\sim 87 \tiny{\%}$ $\sigma_Y$ & 36.70 & 10.00 & 2.69 & 1.00 & 1.00
\end{tabular}
\caption{Bandgap opening frequency and optimal design parameters of the representative unit cell for different maximum permissible stress values $\overline{\sigma_j}$. \label{t:opt_arch}}
\end{table}

\section{Conclusions}
In this work, we investigate the dynamics and the filtering properties of an inertial amplified structure with opposite chirality, whose unit block consists of a hollow-square plate connected to a ring through arch-like ligaments. The geometry and orientation of the ligaments allow to couple the axial and torsional motions of the structure, thus increasing the effective inertia of the system. Both the analytical and numerical dispersion curves reveal a wide bandgap, whose opening frequency is controlled by the fundamental coupled axial-torsional mode of the building block. The transmission spectrum of the finite-sized structure exhibits a wide signal reduction within the bandgap region identified by the dispersion analysis, with vibration amplitudes decreasing of almost two orders of magnitude when two unit cells are considered.

In setting up an experimental prototype, we fabricate a finite-sized lattice composed of two unit cells by selective laser sintering in Nylon 12 and we experimentally calculate the wave transmission through the structure, which supports validation of the analytical and numerical findings.

We complete our study by developing and solving a stress-constrained optimization model to determine the optimal design parameters of the structure that minimize the bandgap opening frequency and, at the same time, fulfill structural requirements. In particular, internal stresses induced by the self-weight of the structure are kept to a low in order to prevent high stress concentrations in critical locations, which can in turn compromise their functionality and structural safety. The external radius of the ring and the cross sectional dimension of the ligament prove to be the geometric variables that most affect the bandgap opening frequency. In particular, to achieve a low frequency bandgap, the external radius of the ring and the cross section dimension of the ligament will assume, when allowed by the stress constraint, the highest and the lowest possible value, respectively.

We believe the inertial amplification mechanism, proposed in this work and enabled by both the arch geometry and the orientation of the ligaments, can be of use for enhancing the design, at various length scales, of vibration mitigation structures and shock-absorbing materials.

\section*{Declaration of Competing Interest}
The authors declare that they have no known competing financial interests or personal relationships that could have appeared to influence the work reported in this paper.

\section*{Acknowledgments}
This work was partially supported by the ETH Research Grant (49 17-1) to E.C.

\newpage

\bibliography{mybibfile}

\newpage
\input{figure}

\end{document}

%% file: figure.tex
\section*{}